\documentclass[11pt]{article}

\usepackage{amsmath}
\usepackage{amsfonts}
\usepackage{amsthm}
\usepackage{amssymb}
\usepackage{amscd}
\usepackage[british]{babel}
\usepackage{graphicx}
\usepackage{psfrag}
\usepackage{epsfig}
\usepackage{rotating}
\usepackage{times}
\usepackage{color}

\theoremstyle{plain}

\newtheorem{remark}{Remark}[section]

\newcommand{\boxend}{\flushright{$\Box$}}

\begin{document}

\title{Viability of the matter bounce scenario in Loop Quantum Cosmology from BICEP2 last data}

\author{Jaume de Haro$^{}$\footnote{E-mail: jaime.haro@upc.edu}
and Jaume Amor\'os$^{}$\footnote{E-mail: jaume.amoros@upc.edu}}

\maketitle

{Departament de Matem\`atica Aplicada I, Universitat
Polit\`ecnica de Catalunya, Diagonal 647, 08028 Barcelona, Spain}

\thispagestyle{empty}

\begin{abstract}
The CMB map provided by the {\it Planck} project constrains the value of the ratio of
tensor-to-scalar perturbations, namely $r$, to be smaller than $0.11$ (95 \% CL).
This bound rules out the
simplest models of inflation. However, recent data from BICEP2 is in strong tension with
this constrain, as it finds a value
 $r=0.20^{+0.07}_{-0.05}$ with $r=0$ disfavored at $7.0 \sigma$, which allows these
simplest inflationary models to survive.
The remarkable fact is that, even though the BICEP2 experiment was conceived to search
for evidence of inflation,
its experimental data matches correctly theoretical results coming from the matter
bounce scenario (the alternative model to the inflationary paradigm).
More precisely, most
bouncing cosmologies do not pass {\it Planck's} constrains due to the smallness
of the value of the tensor/scalar ratio $r\leq 0.11$, but with new
BICEP2 data some of them  fit well with experimental data.
This is the case with the matter bounce scenario in the teleparallel
version of Loop Quantum Cosmology.
\end{abstract}

PACS NUMBERS:{04.80.Cc, 98.80.Bp, 98.80.Qc, 04.60.Pp}


\section{Introduction}
The latest {\it Planck} temperature data for cosmic inflation
constrains the spectral index for scalar perturbations to be $n_s=0.9603\pm 0.0073$,
ruling out exact scale invariance with over $5\sigma$ confidence,
and establishes an upper bound for tensor/scalar ratio given by $r\leq 0.11$
(95 \% CL)  \cite{Ade}. Such data
 shrinks the set of allowed simplest inflationary models: power law potentials
in chaotic inflation {\cite{linde}},
exponential potential models {\cite{Lucchin}, inverse power law potentials \cite{Barrow},
are disfavored because they
 do not provide a good fit to
{\it Planck's} data \cite{Ade, Steinhardt}. In fact, this data set prefers
a subclass of inflationary models with {\it plateau-like} inflation potentials
(see for example \cite{Olive}) and $R^2$ gravity
\cite{Odintsov}.

On the other hand, recent results from the BICEP2 experiment \cite{Kovac}, designed
to look for the signal of gravitational waves in the $B$-mode
power spectrum, lead to the same constrain for the spectral index,
but constrain the
ratio of tensor-to-scalar perturbations
to be $r=0.20^{+0.07}_{-0.05}$ with $r=0$ disfavored at $7.0 \sigma$
(see figure $13$   of \cite{Kovac} to compare {\it Planck's} with BICEP2 data).
This higher value of $r$ extends the set of compatible inflationary models,
allowing back some of the simplest inflationary models cited above.

Dealing with the matter bounce scenario, the alternative to the inflationary paradigm
(see \cite{brandenberger} for a report about bouncing
cosmologies), one encounters a similar problem when one tries to match
{\it Planck's} data with theoretical results: theoretical results
provide, in general, values of $r$ higher than $0.11$ and, then,
to sort out this problem some very complicated mechanism has to be introduced
to enhance the power spectrum
of scalar perturbations \cite{Cai}, reducing the ratio $r$ enough to achieve
the bound $0.11$.
However,  {in this work we will show that} the higher value of $r$ provided by BICEP2 allows the viability
of some  bouncing models. {This is the main goal of the paper.}

{As a matter of fact, we will deal with the matter bounce scenario in Loop Quantum Cosmology (LQC) which, when one only takes into account holonomy corrections,
provides the simplest bounce. More precisely,
it is well known that LQC contains two kind of corrections: holonomy corrections and inverse-volume effects. When one deals with the flat
Friedmann-Lema{\^\i}tre-Robertson-Walker (FLRW) geometry, holonomy corrections always lead to a big bounce (see for instance \cite{singh}), however this could not
 happen when one  considers inverse-volume effects. For example,
when the universe is filled by a field under the action of a non-negative
potential (to guarantee a positive energy density), one will obtain a non bouncing universe
because the Hubble parameter never vanishes (see equations (5) and (8) of \cite{Bojowald}). That is the reason why, in this paper, we will do not take into account
inverse-volume corrections.}

{
On the other hand, for the flat FLRW geometry, it has been recently showed in \cite{Haro1} that holonomy corrected LQC can be formulated
as a particular example of teleparallel $F(T)$ gravity, where $T$ is the so-called {\it torsion scalar} whose value in the flat FLRW spacetime is equal to $-6H^2$.
This new formulation of LQC with holonomy corrections has been named {\it teleparallel LQC}
and only coincides with the standard holonomy corrected LQC in the FLRW geometry. Dealing with cosmological perturbations
both formulations lead to different perturbation equations and, of course, to different results. The reason of this difference is that
in holonomy corrected LQC, working in the Hamiltonian framework, the corresponding perturbation equations are obtained replacing the Ashtekar connection by a suitable sinus function
in the classical Hamiltonian and inserting in  it counter-terms to preserve the algebra of constrains
\cite{Barrau,Cailleteau}). In constrast to holonomy corrected LQC, the perturbation equations in teleparallel LQC are directly obtained, in the Lagrangian framework, from the well-known perturbation equations in teleparallel $F(T)$ gravity \cite{Cai1}. In fact, it has been shown in \cite{Haro} that
for scalar perturbations both formulations lead to the same kind of results, the difference appears when one deals with tensor perturbations,
because in teleparallel LQC the equation of perturbations \cite{Haro} is a {\it regular} equation, but in holonomy corrected LQC the corresponding equation \cite{Cailleteau} has two singular points (at the beginning and end of the super-inflationary phase). This difference
is what leads to completely different results.}

{To show that, we deal with the matter bounce scenario in LQC, where the universe is filled by only a scalar  field  whose potential is the simplest one
leading, at early times, to a matter domination in the contracting phase.
In this case, the conservation equation   is a second order differential equation (a Klein-Gordon equation). Each orbit, i.e. each solution
of this differential equation, depicts a different  matter dominated universe at early times. We will see that  one of these orbits can be calculated analytically (the orbit
that depicts a matter dominated universe for all time), but the other ones
have to be calculated numerically. Then, for all of these orbits
}
we will calculate {analytically and numerically},  the { corresponding}
tensor/scalar ratio {for adiabatic perturbations (we only considers one matter field, meaning there are not entropy perturbations)} coming from  holonomy corrected
and teleparallel LQC,
 and we will check that in the case of teleparallel LQC there are {orbits}
leading to theoretical results that
match correctly with BICEP2 data, and there are other  {orbits}  that provide theoretical results that fit well
with {\it Planck's} data. {On the other hand, we will also show numerically that
 holonomy corrected LQC,   provides theoretical results,  that only   match correctly with {\it Planck's} data.}

The units used in the {paper} are $\hbar=c=8\pi G=1$.

\section{ Constrains on inflationary models from experimental data}
Slow-roll inflation is essentially based in two parameters \cite{Stewart}:
\begin{eqnarray}\label{a1}
\epsilon=-\frac{\dot{H}}{H^2}\quad \mbox{ and }\quad  \eta=2\epsilon-\frac{\dot{\epsilon}}{2H\epsilon},
\end{eqnarray}
{where $\dot{}$ is the derivative with respect to the cosmic time.}

In the slow-roll phase, i.e., when the dynamics of the system is given by equations
\begin{eqnarray}\label{a2}
H^2\cong\frac{V(\bar\varphi)}{3}\quad \mbox{ and }\quad  3H\dot{\bar\varphi}+V_{\bar\varphi}\cong 0,
\end{eqnarray} where $\bar\varphi(t)$ is the homogeneous part of the scalar field,
are given by
\begin{eqnarray}\label{a3}
\epsilon\cong\frac{1}{2}\left(\frac{V_{\bar\varphi}}{V} \right)^2\quad \mbox{ and }\quad  \eta\cong\frac{V_{\bar\varphi\bar\varphi}}{V}.
\end{eqnarray}

Using slow-roll parameters $\epsilon$ and $\eta$ the spectral index for
scalar perturbations and the ratio of tensor-to-scalar perturbations
are given by
\begin{eqnarray}\label{a4}
n_s\cong 1+2\eta-6\epsilon\quad \mbox{ and }\quad  r\cong 16\epsilon.
\end{eqnarray}

To compare theoretical results with current observations we need the number of
e-folds during inflation, namely $N$, which in slow-roll
approximation is given by
\begin{eqnarray}\label{a5}
N=\int_{t_b}^{t_e}Hdt\cong \int_{\bar\varphi_e}^{\bar\varphi_b}\frac{V}{V_{\bar\varphi}}d\bar\varphi,
\end{eqnarray}
where the sub-index $b$ (resp. $e$) refers to the beginning (resp. end) of inflation.

As a first example to compare theoretical with experimental results,
we choose a  power law potential {$V(\bar\varphi)=\lambda \bar\varphi^{2n}$}.
For this potential one has
\begin{eqnarray}\label{a6}
{n_s\cong 1-\frac{4n(n+1)}{\bar\varphi_b^2}, r\cong \frac{32n^2}{\bar\varphi_b^2} \mbox{ and } N\cong \frac{\bar\varphi_b^2-2n^2}{4n},}
\end{eqnarray}
where we have chosen as the end of inflation the condition $\epsilon=1$,
which is equivalent to {$\bar\varphi_e^2={2n^2}$}, and to calculate $n_s$ and
$r$ we have evaluated $\epsilon$ and $\eta$ at the beginning of inflation.

Removing $\bar\varphi_b^2$ in (\ref{a6}), i.e., writing $n_s$ and $r$ in terms of the
number of e-folds, one gets
\begin{eqnarray}\label{a7}
{n_s\cong 1-\frac{2(n+1)}{2N+n}, r\cong \frac{16n}{2N+n} \Longrightarrow n_s\cong 1-\frac{n+1}{8n}r.}
\end{eqnarray}

In the case of a quadratic potential {$n=1$}, for $60$ e-folds,
the minimum needed to solve the horizon and flatness problems if inflation starts at GUT energies \cite{Guth}, one  gets
 $n_s=0.9669$ and $r=0.132$.
When one increases the number of e-folds, $n_s$ increases and $r$ decreases.
Then, for the maximal allowed value of the spectral index  $n_s=0.9676$ one has $r=0.1296$,
which means that
the model with a quadratic potential does not fit well neither with {\it Planck's} nor with
BICEP2 data.

{In the same way,
for the maximum value allowed of the spectral index, i.e. for $n_s=0.9676$ the value of $r$ is minimum and is given by
$r=\frac{8n}{n+1}\times 0.0324$.
Since $r$ increases as long as the parameter $n$ increases, and  its minimum value is $r=0.1296$ (reached when $n=1$),
one can conclude that inflationary power law models are disfavored by ${\it Planck's}$ data.}

However, using BICEP2 data, the model {$n=2$} with $70$ e-folds is acceptable
because it satisfies $n_s=0.9577$ and $r=0.2253$.
To be more specific, from the third equation of (\ref{a7}) $r$ is constrained
to belong in the interval
\begin{eqnarray}\label{interval}{\left(\frac{8n}{n+1}\times 0.0324,\frac{8n}{n+1}\times 0.047\right).}\end{eqnarray}

{
Then,
for $n\geq 1$ the interval (\ref{interval}) has a non-empty intersection with
 $(0.15,0.27)$. This means that for all values of $n\geq 1$, there exist values of $N$ such that
 $n_s$ and $r$ are allowed from BICEP2 data. However, we need that $N$ was greater than $60$, which can be checked as follows:
 First of all, we have
 \begin{enumerate}
  \item For $n=1$, the allowed values of $r$ belong in $\left(0.15, 0.188\right)$.
  \item For $n=2$ the allowed values of $r$ belong in $\left( \frac{8n}{n+1}\times 0.0324 , \frac{8n}{n+1}\times 0.047\right)$.
  \item For $n\geq 3$ the allowed values of $r$ belong in $\left( \frac{8n}{n+1}\times 0.0324 , 0.27\right)$.
 \end{enumerate}}

 {
Finally, from the value of $r$ (the second equation of (\ref{a7})) one has
\begin{enumerate}
 \item For $n=1$,  $N$ belongs in $(42.05,52.8)$.
 \item For $n\geq 2$, one has $N\geq 62,82$,
\end{enumerate}
meaning that for $n\geq 2$, the model matches correctly with BICEP2 data.
}

\vspace{0.5cm}

As a second example we consider $R^2$ gravity, {sometimes called Starobinsky model (see \cite{Haro2} for a detailed description of the model)}.
{In $R^2$ gravity}
 one has \cite{Odintsov}
\begin{eqnarray}\label{R2}
 n_s=1-\frac{2}{N}, r=\frac{12}{N^2}\Longrightarrow n_s=1-\sqrt{\frac{r}{3}}.
\end{eqnarray}

Using the  data $n_s=0.9603\pm 0.0073$ and equation (\ref{R2})
one obtains the constrain $$0.0031\leq r\leq 0.0066,$$
what means that BICEP2 data disregards this model. However,
the model matches correctly with {\it Planck's} data.
Effectively, for $60$ e-folds one has
$n_s=0.9666$ and $r=0,0033$ which enters perfectly in
the range of values obtained from {\it Planck's} temperature anisotropy mesurements.

\section{Calculation of the power spectrum in LQC}

In this section we will obtain  the formulas  to calculate the power spectrum for scalar and tensor
perturbations, in both holonomy corrected and teleparallel LQC,
when one deals with the matter bounce scenario.

{It is well known that, when one only takes into account holonomy corrections,  the modified Friedmann equation in  the flat FLRW geometry is given by the following ellipse in the plane $(H,\rho)$
\begin{eqnarray}\label{Friedmann}
 H^2=\frac{\rho}{3}\left(1-\frac{\rho}{\rho_c}\right),
\end{eqnarray}
where $\rho_c$ is the so-called {\it critical density}.}

{On the other hand, as we have already explained in the introduction, the equation (\ref{Friedmann}) could be obtained as a particular case of
 teleparallel $F(T)$ gravity. In \cite{Haro1} this example has been found to be
 \begin{eqnarray}\label{12}
 F_{\pm}(T)=\pm\sqrt{-\frac{T\rho_c}{2}}\arcsin\left(\sqrt{-\frac{2T}{\rho_c}}\right)+G_{\pm}(T),
\end{eqnarray}
with
\begin{eqnarray}\label{8}
G_{\pm}(T)=\frac{\rho_c}{2}\left(1\pm\sqrt{1+\frac{2T}{\rho_c}}  \right),\end{eqnarray}
where $+$ correspond to the super-inflationary phase, i.e. to $\rho>\rho_c/2$, and $-$ to $\rho<\rho_c/2$.}

{Now, dealing with adiabatic cosmological perturbations in the longitudinal gauge $ds^2=(1+2\Phi)dt^2-a^2(1-2\Phi)d{\bf x}^2$ where $\Phi$ is the Bardeen potential, and 
assuming that
the matter part of the Lagrangian is depicted by only one  scalar field $\varphi=\bar{\varphi}+\delta \varphi$, where $\bar\varphi$ is the
homogeneous part of the field, one can show that
the Mukhanov-Sasaki (M-K) equations for adiabatic perturbations are
 given by\cite{Barrau, Cailleteau, Haro}}

{
\begin{eqnarray}\label{A1}
 v_{S(T); h(t)}''-{c}^2_{s ; h(t)}\Delta v_{S(T); h(t)}-\frac{z_{S(T); h(t)}''}{z_{S(T); h(t)}}v_{S(T); h(t)}=0,
\end{eqnarray}}
where $'$ represents the derivative with respect the conformal time, $S$ means scalar perturbations, $T$ tensor perturbations, $h$ holonomy corrected LQC and $t$
teleparallel LQC,
and the square of the velocity of sound in the corresponding approach is given by
\begin{eqnarray}\label{A3}
c^2_{s,h}\equiv \Omega=1-\frac{2\rho}{\rho_c};\quad c_{s,t}^2=|c_{s,h}^2|
\frac{\arcsin\left(2\sqrt{\frac{{3}}{\rho_c}}H\right)}{2\sqrt{\frac{3}{\rho_c}}H}.
\end{eqnarray}

 Moreover the
M-K variables $ z_{S(T); h(t)}$  and $ v_{S(T); h(t)}$ are defined as follows:
\begin{eqnarray}\label{A2}
 \quad
 z_{S;h}=\frac{a\dot{\bar{\varphi}}}{{H}},\quad  z_{T;h}=\frac{a}{c_{s;h}},\quad z_{S;t}=\frac{a|c_{s;h}|\dot{\bar{\varphi}}}{c_{s;t}{H}},
 \quad z_{T;t}=\frac{a c_{s;t}}{|c_{s;h}|},
\end{eqnarray}
and $v_{S(T); h(t)}=\zeta_{S(T); h(t)}z_{S(T); h(t)}$,
where $\zeta_{S;h(t)}\equiv \Phi+\frac{H}{\dot{\bar\varphi}}\delta\varphi$ is the curvature fluctuation in co-moving coordinates and $\zeta_{T;h(t)}$
is the amplitude of tensor perturbations.

\begin{remark}
From the definitions of the M-S variables we can see that for scalar perturbations, the equations in holonomy corrected and teleparallel LQC are essentially the same. They are singular
at the bouncing point (when $H$ vanishes), and differ with the value of square of the velocity of sound, which in the case of holonomy corrected LQC becomes negative in the
super-inflationary phase ($\rho>\rho_c/2$), but as we will see, to calculate the power spectrum of perturbations
the term containing the Laplacian could be disregarded. In constrast, for tensor perturbations the equations are completely different. In the case of holonomy corrected LQC it
contains two singular points, at the beginning and end of the super-inflationary phase, i.e., when $\rho=\rho_c/2$. This does not happen in the teleparallel version where the corresponding
M-S equation is always regular. We will see that due to this difference the ratio of tensor to scalar perturbations is completly different depending on the approach used.
\end{remark}

Once we have the perturbation equations, we can deal with the matter bounce scenario. In this scenario, in order to have a scale invariant
spectrum, the universe has to be matter dominated, at early times, in the contracting phase. This is due to the duality, pointed out in
\cite{wands}, between matter domination in the contracting phase and de Sitter  regime in the expanding one. Then,  since at early times the holonomy effects can be disregarded
because $\rho\ll \rho_c$ (the universe is in the bottom of the ellipse (\ref{Friedmann})),
and   the universe is matter dominated at this epoch, one will obtain
\begin{eqnarray}\label{A4}
 z_{S;h}=  z_{S;t}=\sqrt{3}a ,\quad  z_{T;h}= z_{T;t}=a,
\end{eqnarray}
where $a(t)=\left(\frac{3}{4}\rho_ct^2+1\right)^{1/3}\cong \left(\frac{3}{4}\rho_c\right)^{1/3}t^{2/3}=\frac{\rho_c}{12}\eta^2$,
being $t$ the cosmic time and $\eta$ the conformal time \cite{Haro}.

As a consequence, at early times, the M-S equations, in Fourier space, will becomes
\begin{eqnarray}
 v_{S(T); h(t)}''+\left(k^2 -\frac{a''}{a}\right)v_{S(T); h(t)}=0\Longleftrightarrow
 v_{S(T); h(t)}''+\left(k^2 -\frac{2}{\eta^2}\right)v_{S(T); h(t)}=0,
\end{eqnarray}
whose solutions are
the  mode functions
\begin{eqnarray}\label{BD}
 v_{S(T); h(t)}=\frac{e^{-ik\eta}}{\sqrt{2k}}\left(1-\frac{i}{k\eta}\right),
\end{eqnarray}
that depict the  Bunch-Davies (adiabatic) vacuum when $\eta\rightarrow-\infty$.

On the other hand, at early times, modes well outside the Hubble radius satisfy  the long wavelength condition $k^2\eta^2\ll 1$, and thus, the M-S equations
(\ref{A1}) can be approximated by
\begin{eqnarray}
 v_{S(T); h(t)}''-\frac{z_{S(T); h(t)}''}{z_{S(T); h(t)}}v_{S(T); h(t)}=0,
\end{eqnarray}
which solution is the so-called {\it long wavelength approximation}
\begin{eqnarray}\label{lwl}
 v_{S(T); h(t)}(\eta)=A_{S(T)}(k)z_{S(T); h(t)}(\eta)+B_{S(T)}(k)z_{S(T); h(t)}(\eta)\int_{-\infty}^{\eta}\frac{d\bar{\eta}}{z_{S(T); h(t)}^2(\bar{\eta})}.
\end{eqnarray}

The long wavelength approximation can be explicitely calculated at early times using  (\ref{A4}), yielding
\begin{eqnarray}\label{A5}
v_{S;h(t)}(\eta)\cong\frac{A_S(k)}{4\sqrt{3}}\rho_c\eta^2-\frac{4B_S(k)}{\sqrt{3}\rho_c}\frac{1}{\eta},\quad
v_{T;h(t)}(\eta)\cong\frac{A_T(k)}{12}\rho_c\eta^2-\frac{4B_T(k)}{\rho_c}\frac{1}{\eta}.
\end{eqnarray}

To obtain the value of these coefficients on has to match, in the long wavelength regime  $k^2\eta^2\ll 1$, the approximate solutions (\ref{A5}) with the exact modes (\ref{BD}),
giving as a result \cite{wilson,Haro}
\begin{eqnarray}\label{21}
 A_S(k)=\frac{A_T(k)}{\sqrt{3}}= -\sqrt{\frac{8}{3}}\frac{k^{3/2}}{\rho_c}, \quad B_S(k)=\sqrt{3}B_T(k)=i\sqrt{\frac{3}{8}}\frac{\rho_c}{2k^{3/2}}.
\end{eqnarray}

{Once we have calculated these coefficients we will use the long wavelength approximation (\ref{lwl}) to calculate, at late times
($\eta\rightarrow \infty$), the curvature fluctuation in co-moving
coordinates $\zeta_{S,h(t)}$ and the amplitude for tensor perturbations $\zeta_{T,h(t)}$, obtaining \cite{Haro}
\begin{eqnarray}
 \zeta_{S(T),h(t)}=\frac{ v_{S(T); h(t)}(\eta)}{z_{S(T); h(t)}(\eta)}= A_{S(T)}(k)+B_{S(T)}(k)R_{S(T); h(t)}\cong B_{S(T)}(k)R_{S(T); h(t)},
\end{eqnarray}
where $R_{S(T),h(t)}\cong\int_{-\infty}^{\infty}\frac{d{\bar\eta}}{z_{S(T),h(t)}^2({\bar\eta})}= \int_{-\infty}^{\infty}\frac{d\bar{t}}{a(t)z_{S(T),h(t)}^2(\bar{t})} $.}

From this result we can calculate the power spectrum of scalar and tensor perturbation, in both approaches, as follows:
\begin{eqnarray}\label{powerspectrum}
 {\mathcal P}_{{S(T);h(t)}}(k) \equiv
 \frac{k^3}{2\pi^2}|\zeta_{S(T);h(t)}|^2
 =\frac{ 3\rho_c^2}{\rho_{pl}}R_{S(T);h(t)}^2,
\end{eqnarray}
where $\rho_{pl}$ is the Planck's energy density, which in our units equals to $64\pi^2$. And also the tensor/scalar ratio of perturbations
\begin{eqnarray}\label{ratio1}
 r_{h(t)}\equiv\frac{{\mathcal P}_{{T;h(t)}}(k) }{{\mathcal P}_{{S;h(t)}}(k) }
 =\frac{R_{T,h(t)}^2}{R_{S,h(t)}^2}.
\end{eqnarray}

To end this Section, two important final remarks are in order:
\begin{enumerate}
 \item
The formulas (\ref{powerspectrum}) and (\ref{ratio1}) are essential to perform numerical and analytic calculation
in the matter bounce scenario. It is also important to note that, in order to obtain them, only a   matter dominated universe at early times in the contracting phase
has been required. Indeed,
in next Section we will provide the simplest example that satisfies this requeriment and allows us to perform, with all the details, all the numerical and analytic calculations.
\item As we have already remarked, the M-S equations (\ref{A1}) contain singular points, which means that there are infinitely many ways to match solutions at these points,
and thus, one has infinitely many mode solutions that lead to infinitely many different power spectrums. However, if one assumes that $\zeta_{S(T);h(t)}(\eta)$ has to be
an analytic function for all time $\eta$, then there is only one solution that satisfies this requirement: the one given by (\ref{lwl}). That is the reason why we use the long wavelength
approximation (\ref{lwl}) to calculate the power spectrum of scalar and tensor perturbations in both approximations.
\end{enumerate}


\section{An specific example}

In this Section we will find a
potential that leads to an analytic solution that depicts, all time, a matter dominated universe.
For this potential we also find numerically all the other solutions and, from formula (\ref{ratio1}), we will calculate, for each solution, their corresponding tensor/scalar ratio.

{To find this potential, first of all, we will solve}
the holonomy corrected Friedmann equation and the conservation
equation for a matter dominated universe (see for instance \cite{singh})
\begin{eqnarray}\label{Friedmann1}
 H^2=\frac{\rho}{3}\left(1-\frac{\rho}{\rho_c}\right);\quad \dot{\rho}=-3H\rho,
\end{eqnarray}
{obtaining} the
following quantities \cite{Haro}
\begin{eqnarray}\label{13}
a(t)=\left(\frac{3}{4}\rho_ct^2+1\right)^{1/3}
\mbox{ and }\quad
\rho(t)=\frac{\rho_c}{\frac{3}{4}\rho_ct^2+1}.
\end{eqnarray}

To find
such potential, one can impose that the pressure vanishes, i.e.,
$P\equiv \frac{\dot{\bar\varphi}^2}{2}-V(\bar\varphi)=0$, which
leads to the equation
\begin{eqnarray}
 \dot{\bar\varphi}^2(t)=\rho(t)\Longleftrightarrow \dot{\bar\varphi}^2(t)=\frac{\rho_c}{\frac{3}{4}\rho_ct^2+1},
\end{eqnarray}
where we have used the second equation of (\ref{13}).

This equation has the particular solution
\begin{eqnarray}\label{sol}
 \bar\varphi(t)=\frac{2}{\sqrt{3}}\ln\left(\sqrt{\frac{3}{4}\rho_c} t+\sqrt{\frac{3}{4}\rho_c t^2+1}  \right),
\end{eqnarray}
which leads to the potential
\begin{eqnarray}\label{pot}
 V({\bar\varphi})=2\rho_c\frac{e^{-\sqrt{3}{\bar\varphi}}}{\left(1+e^{-\sqrt{3}{\bar\varphi}}\right)^2}.
\end{eqnarray}

It is important to realize that {the analytic}
 solution (\ref{sol}) is special in the sense that it satisfies for all time $\dot{\bar\varphi}^2(t)/2=V(\bar\varphi(t))$, that is,
 if the universe is described by this solution, it will be matter  dominated all the time. However, {all} the other
solutions, that is, the {other} solutions {that only can be obtained numerically from} the  conservation equation
\begin{eqnarray}\label{KG} \dot{\rho}=-3H_{\pm}(\rho+P)\Longleftrightarrow
 \ddot{\bar{\varphi}}+3{H}_\pm\dot{\bar{\varphi}}+V_{\bar\varphi} =0,
\end{eqnarray}
where the Hubble parameter is equal to ${H}_- =-\sqrt{\frac{{\rho}}{3}(1-\frac{\rho}{\rho_c})}$ in the contracting phase  and
${H}_+ =\sqrt{\frac{{\rho}}{3}(1-\frac{\rho}{\rho_c})}$ in the expanding one,
do not lead to a matter-dominated universe {all the time}. Only at early and
 late times the universe is matter dominated because the solution (\ref{sol}) is
a global repeller at early times and
a global attractor at late ones {(see \cite{ha} for a demonstration)}.

Once we have introduced the simplest potential for the matter bounce scenario in LQC,
we deal with {scalar} perturbations.
In the case of
 holonomy corrected LQC
 for the analytic solution
(\ref{sol})
  { one has $z_{S;h}=\frac{2a^{5/2}(t)}{\sqrt{\rho_c}t}$} \cite{wilson}, and which leads,
after using formula (\ref{powerspectrum}), to
\begin{eqnarray}\label{25}
 {\mathcal P}_{S;h}(k)=\frac{\pi^2}{9}\frac{\rho_c}{\rho_{pl}}.
\end{eqnarray}

On the other hand, in teleparallel LQC, whose perturbation equations, {as we have explained in the introduction}, are the ones
of $F(T)$ gravity \cite{Cai1} applied to a model (see eq. (2.12) and (2.23) of \cite{Haro})
whose teleparallel Friedmann equation
coincides with the holonomy corrected one (\ref{Friedmann}),
for the particular solution (\ref{sol}) one has
{\begin{eqnarray}\label{26}
 z_{S;t}(t)=2\left(\frac{3}{\rho_c}\right)^{1/4}\frac{a(t)|t|^{1/2}}{t\sqrt{\arcsin\left(\frac{\sqrt{3\rho_c}|t|}{a^3(t)} \right)}},
\end{eqnarray}}
giving as a power spectrum
{
\begin{eqnarray}\label{27}
 {\mathcal P}_{S;t}(k)
=\frac{16}{9}\frac{\rho_c}{\rho_{pl}}{\mathcal C }^2,
\end{eqnarray}}
where ${\mathcal C}\cong 0.9159$ is Catalan's constant.

This result has to be compared with the seven-year data of WMAP \cite{Larson},
which constrains the value of the power spectrum for
scalar perturbations to be ${\mathcal P}(k)\cong 2\times 10^{-9}$, which means that,
in both cases (holonomy corrected and teleparallel LQC),
when one considers the solution (\ref{sol}),
the value of the critical density has to be of the order $\rho_c\sim 10^{-9}\rho_{pl}$.

{ Dealing with the tensor/scalar ratio of perturbation,}
for the analytical solution (\ref{sol}), in holonomy corrected LQC, {after using formula (\ref{ratio1})} one has {$r_h=0$}
which is an abnormally small value, and in teleparallel LQC
we have obtained the following very high value {$r_t= 3\left(\frac{Si(\pi/2)}{{\mathcal C}}\right)^2\cong 6.7187$}, where $Si(x)\equiv\int_0^x\frac{\sin y}{y}dy$
is the Sine integral function.

However, these results do not mean that the matter bounce {model depicted by the potential (\ref{pot})} has to be disregarded. What they mean is that, for orbits (solutions
of (\ref{KG})) near the solution (\ref{sol}), the theoretical results {given by holonomy corrected and teleparallel LQC} do not match
with the current experimental data. But, {as we will see numerically, in the case of teleparallel LQC}, there are  other orbits
{whose theoretical results} fit well with data obtained from {\it Planck}, {and others whose
theoretical results, match with} BICEP2 data.
{And, in the case of holonomy corrected LQC, we will also
show that all the orbits satisfy {\it Planck's} constrain $r\leq 0.11$.}

Dealing with the spectral index, the matter bounce scenario provides a  power spectrum
exactly scale invariant, i.e., $n_s=1$  not agreeing
 with current data
{ $n_s=0.9603\pm 0.0073$ \cite{Ade}}, which states that is nearly scale invariant with a slight red tilt.
The problem is easily solved
if one assume that at early times, in the contracting phase, the universe has an state equation
of the form $P=\omega\rho$ with
$|\omega|\leq 1$. In LQC a potential that leads to this kind of universe is \cite{wilson}
 \begin{eqnarray}\label{pot1}
 V({\bar\varphi})=2\rho_c(1-\omega)\frac{e^{-\sqrt{3(1+\omega)}{\bar\varphi}}}{\left(1+e^{-\sqrt{3(1+\omega)}{\bar\varphi}}\right)^2}.
\end{eqnarray}

In fact, this potential provides an analytic orbit (an analytic solution of (\ref{KG}))
that depicts an universe whose equation of state is  $P=\omega\rho$ all the time. Moreover,
at early times this orbit is a repeller and at late times an attractor, meaning that all
the orbits represent a universe that at early
and late times has as  equation of state $P=\omega \rho$.
As a consequence, for all the orbits of the system the spectral index is given by \cite{wilson}
$n_s=1+12\omega$. Then, to
match with observational data one only has to choose $\omega=-0.0033\pm 0.0006$.

{It is important to realize that there are other potentials whose orbits, at early times,
depict a universe with equation of state $P=\omega \rho$. In fact, as we have recently  showed in \cite{ha},
 all potentials with the asymptotic   form $V({\bar\varphi})\sim \rho_c e^{-\sqrt{3(1+\omega)}|\bar\varphi|}$ when $|{\bar\varphi}|\rightarrow \infty$,
 have this property.} 

Finally, is {also} important to stress  that for these small values of $\omega$
the corresponding formulae for the power spectrum and the
tensor/scalar ratio  do not change significatively, i.e., we can continue using
 {formulas (\ref{powerspectrum}) and
(\ref{ratio1}).}

\subsection{Numerical results}

Our numerical study is based in the numerical resolution of equation (\ref{KG}) {for the potential (\ref{pot})} .
To perform this calculation, one has to take into account that
in LQC the orbits start at early times in the contracting phase ($H<0$),
and when its energy density reaches the critical density $\rho=\rho_c$ the universe
bounces and enters in the expanding phase ($H>0$). Then, to obtain the phase portrait
of the system in the plane $(\bar\varphi,\dot{\bar\varphi})$, for any initial
 condition $(\bar\varphi_0,\dot{\bar\varphi}_0)$ one has to integrate numerically
equation (\ref{KG}) with $H={H}_-$
forward in time, and when the orbit reaches the curve $\rho=\rho_c$ at some point
$(\bar\varphi_1,\dot{\bar\varphi}_1)$, one has to integrate
numerically forward in time equation (\ref{KG}) with $H={H}_+$ for the new
initial condition $(\bar\varphi_1,\dot{\bar\varphi}_1)$.
The phase portrait is pictured in figure $1$.

\begin{figure}[h]
\begin{center}
\includegraphics[scale=0.3]{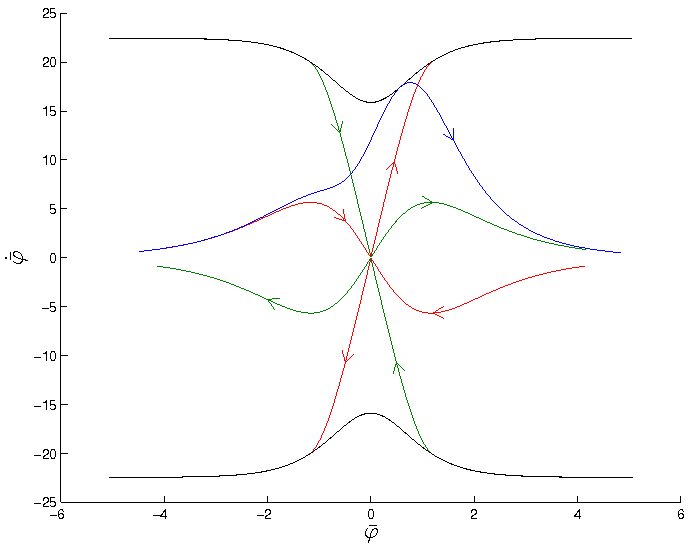}

\end{center}

\caption{{\protect\small Phase portrait: the
black curves are defined by $\rho=\rho_c$, and depict the points where the universe bounces.
The point $(0,0)$ is a saddle point, red (resp. green) curves are the invariant
curves  in the contracting (resp. expanding) phase. The blue curve corresponds
to an orbit different from the analytically one (\ref{sol}).
Note that, before (resp. after) the bounce the blue curve does not cut the red
(resp. green) curves. It is important to realize that the allowed orbits are
those that catch the
black curve in the region delimited by an unstable red curve and an stable green curve,
because for  orbits that do not satisfy this condition, $\dot{\bar\varphi}$ vanishes
at some time,
meaning that its corresponding power spectrum diverges.
}}
\end{figure}

For a wide range of the orbits calculated numerically, we have obtained for the power spectrum of scalar perturbations,
which, in  the case of potential (\ref{pot}), is proportional to the ratio $\rho_c/\rho_{pl}$
for all the orbits of the system (\ref{KG}), the following results:

\begin{enumerate}
 \item In holonomy corrected LQC, the minimum value of  {${\mathcal P}_{S;h}(k)$} is
obtained for the orbit that at  bouncing time satisfies $\bar\varphi\cong -0.9870$, for that
 orbit we have obtained {${\mathcal P}_{S;h}(k)\cong 23\times 10^{-3}\frac{\rho_c}{\rho_{pl}}$}.
 \item In teleparallel LQC the orbit which gives the minimum value of the power spectrum
satisfies, at bouncing time, $\bar\varphi\cong -0.9892$ and the value of the power spectrum is
 approximately the same as in holonomy corrected LQC
 {${\mathcal P}_{S;t}(k)\cong 40\times 10^{-3}\frac{\rho_c}{\rho_{pl}}$}.
\end{enumerate}

Then for those orbits, in order to match with the current experimental result
${\mathcal P}_S(k)\cong 2\times 10^{-9}$, in both theories one has to choose
$\rho_c\sim 10^{-7}\rho_{pl}$
which
is
$2$ orders greater than the value needed using the analytic solution. This result,
{as was pointed out in \cite{wilson}, is in tension with the current value of the critical density $\rho_c\sim 0.4 \rho_{pl}$, obtained relating the black hole entropy in 
Loop Quantum Gravity (LQG)
with the Bekenstein-Hawking entropy formula
\cite{meissner}. To solve this discrepancy  in holonomy corrected LQC, in \cite{wilson1} the matter bounce scenario has been improved introducing 
a sudden transition between the matter-domination period and
an ekpyrotic phase. Then, in this new matter-ekpyrotic bouncing scenario,  it is heuristically argued in \cite{wilson1} that
the power spectrum of scalar perturbations is entirely determined by the value of the Hubble parameter at the beginning
of the ekpyrotic phase, and thus, the WMAP data ${\mathcal P}(k)\cong 2\times 10^{-9}$, does not fix the value of the critical density, rather it determines the value of the
Hubble parameter at the beginning of the ekpyrotic phase.
In constrast,
this tension  does not affect teleparallel LQC, because teleparallel LQC is a teleparallel  $F(T)$ example that mimics, only in the flat FLRW geometry, holonomy corrected LQC.
Thus, results comming from LQG are not related with teleparalell LQC,
where $\rho_c$ is
merely a parameter whose value has to be obtained from observations.}

We have also calculated the ratio of tensor-to-scalar perturbations, which is independent
on the parameter $\rho_c$, for the potential (\ref{pot})
 in teleparallel LQC using formula {(\ref{ratio1})}. Its value in admissible solutions
(those with $\dot{\bar\varphi} \neq 0$ at all times) ranges continuously from a minimal
value {$r_t=0$}, attained by the orbit with the universe bouncing at {$\bar\varphi\cong -1.205$} and $\bar\varphi\cong 1.205$,
to the maximal value {$r_t\cong 6.7187$}, attained by the solution (\ref{sol}) bouncing
at $\bar\varphi =0$. The confidence interval $r =0.20^{+0.07}_{-0.05}$
derived from BICEP2 data is realized by solutions bouncing when
{$\bar\varphi \in [-1.162,-1.144]\cup[1.162,1.205]$}, and the bound $r\leq 0.11$ provided by {\it Planck's} experiment is realized by
solutions bouncing when {$\bar\varphi \in [-1.205,-1.17]\cup[1.17,1.205]$}.
Moreover, subtracting various dust models the tensor/scalar ratio in BICEP2 experiment could be shifted to $r=0.16^{+0.06}_{-0.05}$ with $r=0$ disfavored
at $5.9\sigma$. Then,  this confidence interval is realized by solutions bouncing when
{$\bar\varphi \in [-1.17,-1.1496]\cup [1.1496,1.17]$}.

{For orbits that match correctly with BICEP2 data we have also calculated the power spectrum of scalar perturbations, obtaining that they belong is 
in the range $2\times 10^{-1} \frac{\rho_c}{\rho_{pl}}\leq {\mathcal P}_{S;t}(k)\leq 36\times 10^{-2} \frac{\rho_c}{\rho_{pl}}$,
meaning that, to match with WMAP data, in teleparallel LQC the value of $\rho_c$ has to be of the order $10^{-8}\rho_{pl}$.}

On the other hand, in holonomy corrected LQC, {when there is only one matter field}, numerical results show that the allowed
 orbits provide values of {$r_h$} in the interval $[0,0.12]$, matching correctly with
{\it Planck's} constrain $r\leq 0.11$, but not with BICEP2 data. {Moreover, if one wants to obtain theoretical results in 
holonomy corrected LQC that fit well with BICEP2 data, it
is argued in \cite{wilson1} that one will have to introduce more than one matter field. Then, entropy perturbations might be important in a matter-ekpyrotic bounce scenario.
But, this is a question that needs further investigation.}

{Finally, in figure 2 we have plotted the graphic of $r_t$ and $r_h$ in function of the bouncing value of the orbit.}
\begin{figure}[h]
\begin{center}
\includegraphics[scale=0.30]{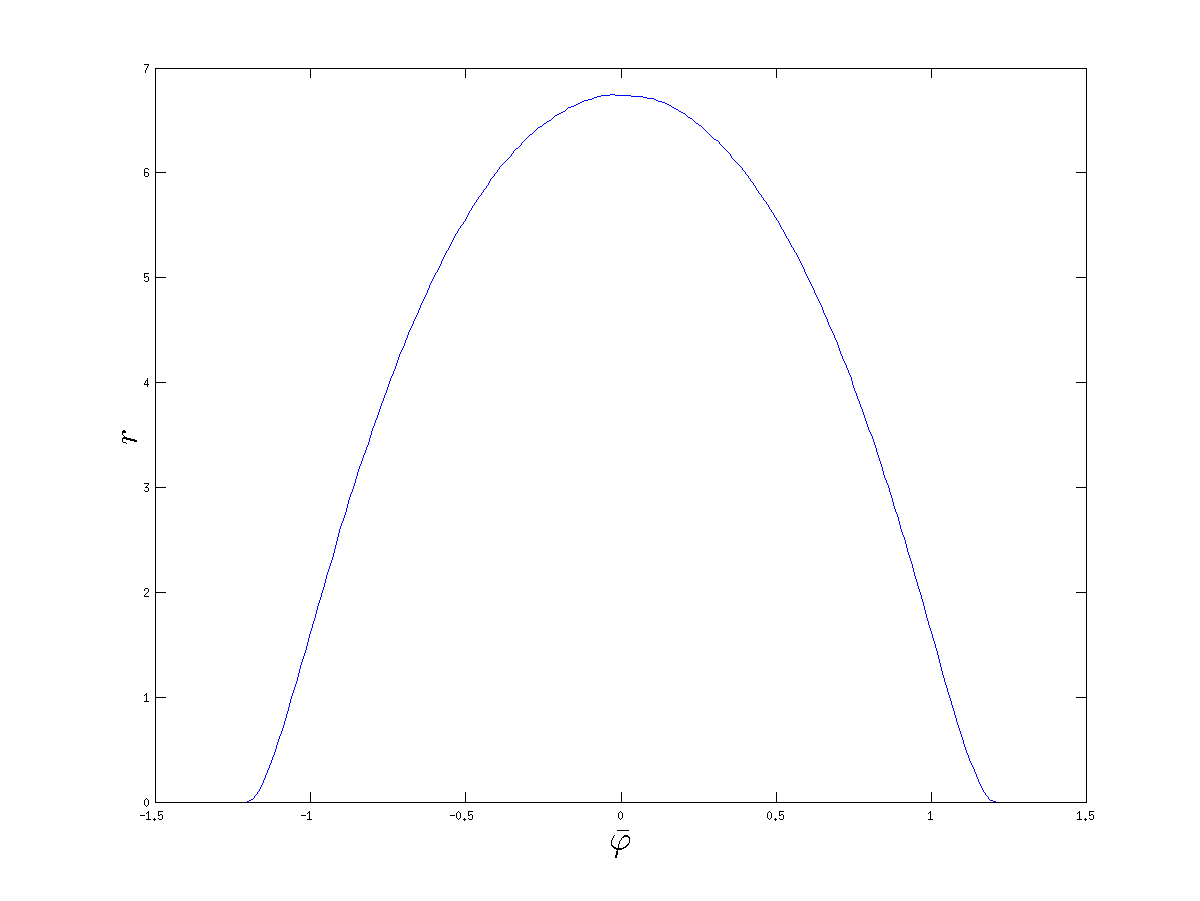}
\includegraphics[scale=0.30]{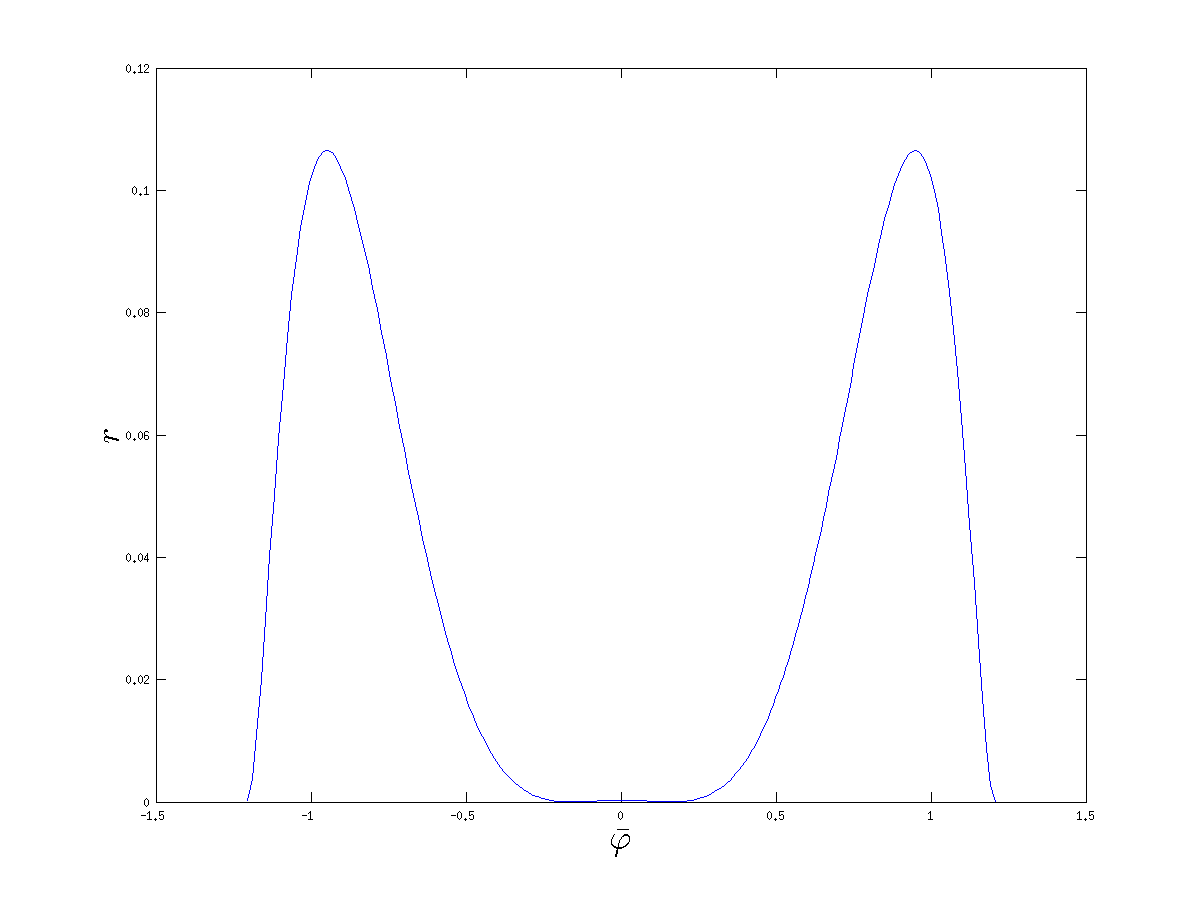}
\end{center}

\caption{{\protect\small {Tensor/scalar ratio for different orbits  in function of the bouncing value of $\bar\varphi$. In the first picture for  teleparallel LQC, and in the second one for
holonomy corrected LQC.}}}
\end{figure}

\section{Conclusions}
In this work we have studied cosmological perturbations produced by one matter scalar field (adiabiatic perturbations) in the context of holonomy corrected and teleparallel 
LQC. We have
explained that, in the flat FLRW geometry, both formulations coincide, but dealing with perturbations they provide different results. This is basically due to tensor perturbations, that
satisfy completely different equations depending on the formulation used. Our results show that holonomy corrected LQC only fits well with {\it Planck's} data due to the low value of
the tensor/scalar ratio provided by this theory. Then, since
{\it Planck's} and BICEP2 data are in strong tension (they provide completely different experimental data for the ratio of tensor to scalar perturbations), holonomy corrected LQC is 
only viable
if the correct experimental data are the ones given by the {\it Planck} project.
 However, teleparallel LQC provides theoretical results that match with {\it Planck's} data and others that fit well with BICEP2. Then, whichever between the {\it Planck} or BICEP2 experimental results are
most accurate (at this moment, there is not an answer), teleparallel LQC always has a set of solutions whose theoretical results match correctly with 
the accurate experimental data.

\vspace{0.5cm}

\centerline{\bf Aknowledgements}

\vspace{0.5cm}

We thank Professor Sergei D. Odintsov for correspondence and useful comments.
This investigation has been
supported in part by MINECO (Spain), project MTM2011-27739-C04-01, MTM2012-38122-C03-01,
 and by AGAUR (Generalitat de Ca\-ta\-lu\-nya),
contracts 2009SGR 345 and 994.



\end{document}